\documentclass[12pt,preprint]{aastex} %Submission

\slugcomment{{\bf In press, Astrophysical Journal Letters}}

\shorttitle{The {\bf SPACE} model}

\shortauthors{Gounelle et al.}

\begin{document}

\title{Supernova Propagation And Cloud Enrichment: \\
A new model for the origin of $^{60}$Fe in the early solar system}

\author{Matthieu Gounelle\altaffilmark{1}, Anders Meibom\altaffilmark{1}, Patrick Hennebelle\altaffilmark{2} \& Shu-ichiro Inutsuka\altaffilmark{3}}

\altaffiltext{1}{Laboratoire d'\'{E}tude la Mati\`ere
Extraterrestre, Mus\'{e}um National d'Histoire Naturelle, 57 rue
Cuvier, 75 005 Paris, France.} \email{gounelle@mnhn.fr}

\altaffiltext{2}{Laboratoire de Radioastronomie Millim\'{e}trique,
\'{E}cole Normale Sup\'{e}rieure et Observatoire de Paris, 24 rue
Lhomond, 75005 Paris, France.}

\altaffiltext{3}{Department of Physics, Kyoto University, Kyoto,
606-8502, Japan.}

\begin{abstract}
The radioactive isotope $^{60}$Fe ($T_{1/2} = 1.5 $ Myr) was present
in the early solar system. It is unlikely that it was injected
directly into the nascent solar system by a single, nearby
supernova. It is proposed instead that it was inherited during the
molecular cloud stage from several supernovae belonging to previous
episodes of star formation. The expected abundance of $^{60}$Fe in
star forming regions is estimated taking into account the
stochasticity of the star-forming process, and it is showed that
many molecular clouds are expected to contain $^{60}$Fe (and
possibly $^{26}$Al [$T_{1/2} = 0.74 $ Myr]) at a level compatible
with that of the nascent solar system. Therefore, no special
explanation is needed to account for our solar system's formation.
\end{abstract}

%In
%our model, which is called the {\bf S}upernova {\bf P}ropagation
%{\bf A}nd {\bf C}loud {\bf E}nrichment ({\bf SPACE}) model,
%$^{60}$Fe is introduced into the molecular cloud by supernovae from
%previous generation of stars. A quantitative stochastic model for


\section{Introduction}
\label{sec-intro}

Short-lived radionuclei (SLRs) are radioactive isotopes with
half-lives shorter than 100 Myr, which were present in the early
solar system (ESS, Russell et al. 2001). Because of their relatively
high abundances with respect to that of the interstellar medium
(ISM), some SLRs must have been produced within, or close in space
and time to the ESS rather than during continuous Galactic
nucleosynthesis (e.g. Meyer \& Clayton 2000).

Iron-60 ($T_{1/2} = 1.5 $ Myr) holds a special position because it
is only produced efficiently by stellar nucleosynthesis unlike other
SLRs, which can also be made in the protoplanetary disk via
irradiation of dust/gas by accelerated energetic particles such as
protons (Lee et al. 1998). As such, $^{60}$Fe provides important
clues about the immediate stellar environment of the nascent solar
system (Montmerle et al. 2006). AGB stars are not considered a
likely source of $^{60}$Fe in the solar system because of their low
probability of encounter with a star-forming region (Kastner \&
Myers 1994).

Elaborating on the pioneering work of Cameron \& Truran (1977), two
different quantitative scenarios with {\it nearby, single} supernova
have been proposed whereby $^{60}$Fe is injected either into the
solar protoplanetary {\it disk} (e.g. Ouellette et al. 2007) or into
the molecular cloud (MC) {\it core} progenitor of our solar system
(e.g. Cameron et al. 1995). Some models envision both possibilities
(Takigawa et al. 2008). In the supernova-disk scenario, the small
size of the disk requires that it lies within 0.4 pc from the
injecting supernova belonging to the same stellar cluster (Ouellette
et al. 2007). However, when massive stars become supernovae after a
few Myr of evolution, remaining disks around low-mass stars are
several pc away from the massive star (e.g. Sicilia-Aguilar et al.
2005), and receive only minute amounts of SLRs (Williams \& Gaidos
2007; Gounelle \& Meibom 2008). In the supernova-core scenario, the
supernova shockwave triggers the core gravitational collapse in
addition to delivering SLRs only for very restricted conditions in
term of distance and shockwave velocity (Boss et al. 2008).

At present, there is therefore no satisfying model which can explain
the presence of $^{60}$Fe in the ESS. Here, we quantitatively
evaluate a scenario proposing that $^{60}$Fe was inherited in the
progenitor molecular cloud from supernov{\it ae} belonging to
previous episodes of star formation. This scenario differs from
scenarios favoring direct injection into a disk/core from a
contemporaneous, single, nearby supernova mainly because the
$^{60}$Fe adduction occurs at the larger MC scale. As previous
stellar models trying to account for the presence of $^{60}$Fe in
the early solar system (Ouellette et al. 2007; Meyer \& Clayton
2000; Mostefaoui et al. 2005, Boss et al. 2008), our model cannot
solve the problem of $^{53}$Mn ($T_{1/2} = 3.7 $ Myr) overproduction
relative to $^{60}$Fe and their relative abundances in the early
solar system (e.g. Wasserburg et al. 2006).
%Our new
%scenario is based on an attractive model for the formation of
%molecular clouds (MCs) which is currently emerging in the
%astronomical community (e.g. Hartmann et al. 2001).
%Compared to previous works (Cameron \& Truran
%1977; Clayton 1983), our model provides an astrophysical setting for
%the origin of $^{60}$Fe in line with the present understanding of
%star formation within molecular clouds.

\section{The {\bf S}upernova {\bf P}ropagation {\bf A}nd {\bf
C}loud {\bf E}nrichment Model}

\label{sec-model}

\subsection{Model sketch}

\label{sec-sketch}

Recently, a new paradigm concerning the formation mechanisms and
lifetimes of molecular clouds emerged (see Hennebelle et al. 2007
and references therein). In this new paradigm, referred to as the
turbulent convergent flow model, MCs result from the collision of
coherent flows and large-scale shocks in the interstellar medium
driven by winds from massive stars and supernova explosions. Such
collisions compress the interstellar atomic gas and after 10 to 20
Myr of evolution, the gas is dense enough to be shielded from the UV
radiation and to become molecular (e.g. Glover \& MacLow 2007). Star
formation follows immediately after the formation of the dense
molecular gas. The turbulent convergent flow model provides a
natural explanation for the wind-swept appearance of molecular
clouds (V\'{a}zquez-Semadeni et al. 2005), but is also consistent
with the short lifetime of MCs (Hartmann et al. 2001) and the
formation of stars in molecular clouds within a crossing time
(Elmegreen 2000). In addition, the turbulent convergent flow model
elegantly accounts for the division of OB associations in subgroups
of different ages (Lada \& Lada 2003).
%Famous examples are
%the Orion region containing the OB1a ($\sim$ 10 Myr), OB1b ($\sim$ 5
%Myr), OB1c ($\sim$ 4 Myr), OB1d ($\sim$ 1 Myr) subregions.
A famous example is the Scorpio-Centaurus region made of the Lower
Centaurus Crux (LCC, $\sim$ 16 Myr), the Upper Centaurus Lupus (UCL,
$\sim$ 17 Myr) and the Upper Scorpius (Upper Sco, $\sim$ 5 Myr)
subregions (Preibisch \& Zinnecker 2007).
%The range in ages in a
%given region has long been interpreted as indicative of sequential
%star formation (e.g. Elmegreen \& Lada 1977) which is here
%reformulated in term of {\it sequential molecular cloud formation}.

If the turbulent convergent flow model is correct, relatively high
concentrations of $^{60}$Fe and other radioactivities with
half-lives $\ga$ 1 Myr are expected in molecular clouds. This is
because supernova ejecta, whose compression effects build molecular
clouds, also carry large amount of radioactive elements such as
$^{60}$Fe. Although it can take as long as 20 Myr to build a
molecular cloud depending on the starting density of the atomic gas,
live $^{60}$Fe is continuously replenished in the second generation
molecular cloud by supernovae originating from the first episode of
star formation, which explode every few Myr. We therefore suggest
that $^{60}$Fe in the ESS was inherited from multiple supernovae
belonging to previous episodes of star formation and name our model
{\bf SPACE} for {\bf S}upernova {\bf P}ropagation {\bf A}nd {\bf
C}loud {\bf E}nrichment.

\subsection{Quantitative estimate}
\label{sec-quant}

%\subsection{{\bf SPACE 1} in the context of the turbulent convergent flow model }

%\label{sec-flow}

%In the {\it turbulent convergent flow} model, molecular clouds
%result from the collision of coherent flows and large-scale shocks
%in the interstellar medium prompted by winds from massive stars and
%supernovae explosions of previous generations of stars (e.g.
%V\'{a}zquez-Semadeni et al. 1995). Such collisions compress the
%interstellar atomic gas. After 10 to 20 Myr of evolution, the gas is
%dense enough to be shielded from the UV radiation and to become
%molecular (e.g. Hartmann et al. 2001). Star formation follows
%immediately the apparition of the dense molecular gas.

\begin{figure}
%\epsscale{1.8}
\plotone{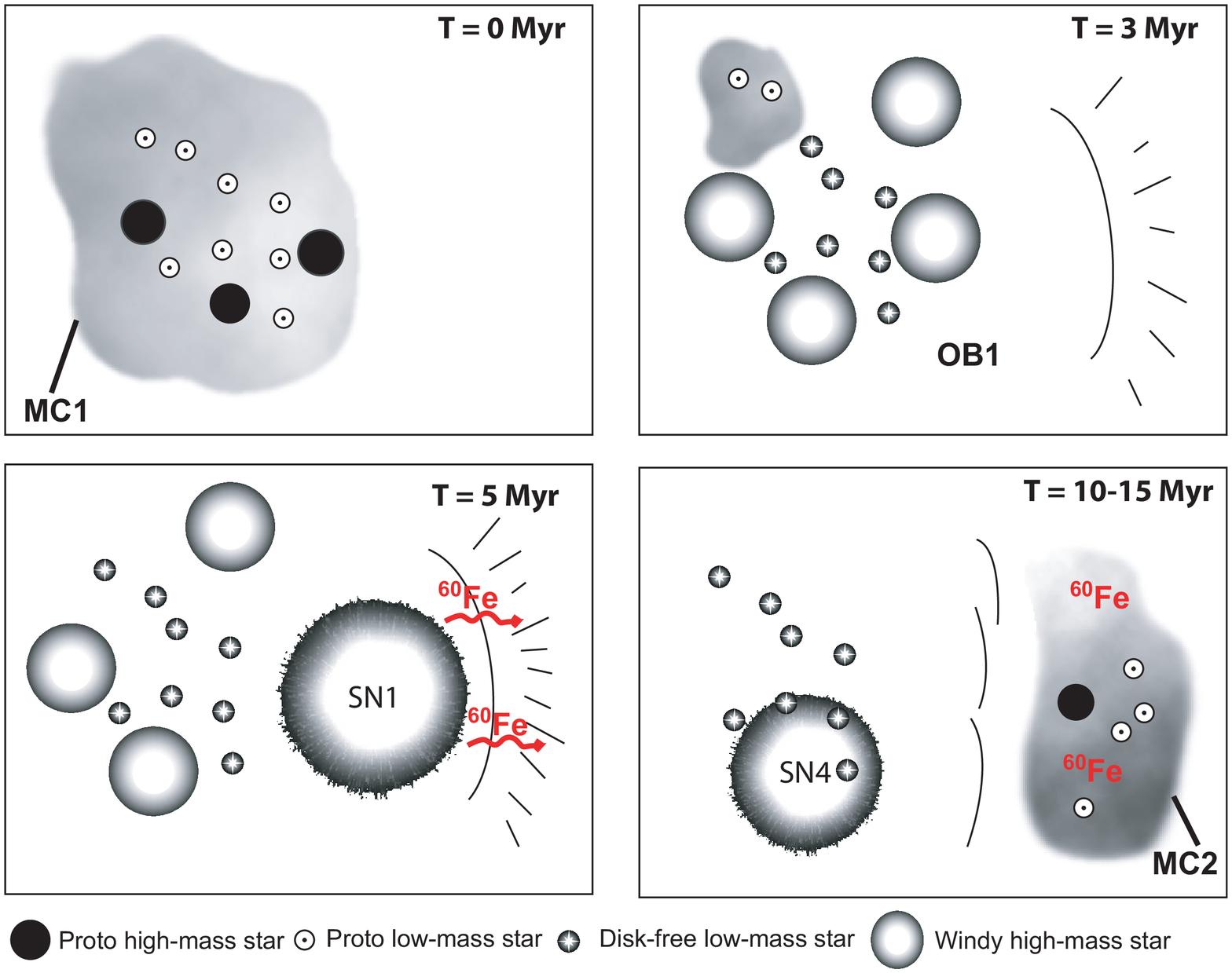} \caption{Sketch of the {\bf SPACE} model. At t = 0
Myr, star formation starts in MC1. At t = 3 Myr, high-mass stars
emit powerful winds which dissipate the molecular gas and start to
accumulate interstellar gas further away. At t = 5 Myr, the first
supernova in OB1 explodes and mixes $^{60}$Fe in the swept-up gas.
At t = 10-15 Myr, the swept-up gas becomes dense enough to become
molecular and star formation starts in MC2. As a result of
sequential molecular cloud formation, MC2 contains a relatively high
abundance of $^{60}$Fe produced by the previous supernovae (4 in
that particular case), which contributed to its formation.}
\label{fig-sketch}
\end{figure}

We consider a first generation of stars formed in molecular cloud 1
(MC1) and a second generation of stars formed in molecular cloud 2
(MC2). After dissipation of the gas, the first (second) generation
of stars become the OB1 (OB2) association. In our model, a number of
supernovae from MC1 deliver $^{60}$Fe into the second generation
molecular cloud MC2 (Fig. \ref{fig-sketch}).
%In what follows, we calculate how much $^{60}$Fe
%is produced by the association OB1 containing N$_1$ stars.

%To fix ideas, we will
%identify MC1/OB1 (MC2/OB2) to the LCC-UCC (Upper Sco) star-forming
%regions.

The mass of $^{60}$Fe in MC2 as a function of time $t$ (time zero
being the onset of star formation in MC1) reads:
\begin{equation} M_{\rm MC2}(^{60}{\rm Fe})[t]  = f \; \eta \; \sum_{i=1}^{i={\rm N_{\rm SN}}} Y_{\rm SN_i}(^{60}{\rm Fe}) e^{-(
t-t_{i})/\tau}, \label{eq-1}\end{equation} where $f$ is a
geometrical dilution factor, $\eta$ is the mixing efficiency,
N$_{\rm SN}$ is the number of supernovae which have exploded in OB1
before time $t$, $\tau$ is the mean life of $^{60}$Fe, $ Y_{\rm
SN_i}(^{60}{\rm Fe})$ is the $^{60}$Fe yield of the i$^{\rm th}$
supernova in MC1 and $t_{ i}$ is the time of the i$^{\rm th}$
supernova explosion in MC1.

%Ouellette et al. (2007) proposed that the mixing efficiency for iron
%atoms from a supernova ejecta into a dense disk is nearly 100 \%. Because
%the proto-molecular cloud matter is far more diffuse than the disk
%matter, mixing will be more efficient and we can reasonably assume
%$\eta = 1$. Turbulence in the proto-molecular cloud will also
%contribute to an efficient mixing between the supernova ejecta and the
%pre-MC matter.

%Given that in the {\bf SPACE 1} model, the supernovae from OB1 face
%by definition MC2 (see Figure 1), we expect $f$ to be close to 0.5.
%We conservatively assume that only one tenth of the supernova ejecta
%contributes to the sweeping-up of atomic gas and the adduction of
%$^{60}$Fe into the new molecular cloud, therefore $f=0.1$ This low
%value of $f$ takes into account the possibility for $^{60}$Fe atoms
%to escape the Galactic disk within Galactic fountains (e.g. Bally et
%al. 2005).

The stellar masses ($M$) in MC1 are calculated following the stellar
Initial Mass Function (IMF), using the generating function of Kroupa
et al. (1993), $ M=0.01+ (0.19 \xi^{1.55}+0.05 \xi^{0.6})/(1-\xi
^{0.58})$, where $\xi$ is a random number to be chosen between 0 and
1 (Brasser et al. 2006). We consider only the distributions whose
most massive star is less massive than 150 M$_{\sun}$, a likely
upper limit for stellar masses (Weidner \& Kroupa 2006).
%This
%function generates a set of stars whose average mass is 0.4
%M$_{\sun}$.
% (compared to 0.51 M$_{\sun}$ for the Scalo IMF, see
%Williams \& McKee 1997), and more
Importantly, we find that $f_{\rm SN}$, the fraction of stars more
massive than 8 M$_{\sun}$ which will go supernova, is 2.3 $\times$
10$^{-3}$.
% (i.e. which
%will go supernova) lies in the middle of the usually considered range of
%1-3 $\times$ 10$^{-3}$ (Higdon \& Lingenfelter 2005; Adams \&
%Laughlin 2001).

%Our value of
%$f_{\rm SN}$ is very close to that obtained with the Scalo IMF
%(Scalo 1986), $f_{\rm SN}$ = 2.6 $\times$ 10$^{-3}$ (McKee \&
%Williams 1997).

The yields of $^{60}$Fe have been determined for a diversity of
supernovae, corresponding to progenitor massive stars with masses
ranging from 11 to 120 M$_{\sun}$ (Woosley \& Weaver 1995; Rauscher
et al. 2002; Limongi \& Chiffi 2006; www.nucleosynthesis.org).
Though in relatively good agreement, the yields somehow vary because
of differences in the stellar and nuclear physics used by the
different groups. For a given mass, when different yields are
available, we use the average of the different yields. For stellar
masses for which the yields are unknown, we take the yield of the
star closest in mass. The explosion time $t_i$ of each massive star
depends on the mass of the progenitor and is given by
%the expression ${\rm log}( t_
%i)={1.4}/{({\rm log }M_ i)^{1.5}}$, (where $M_i$ is the mass of star
%i) deduced by Williams \& Gaidos (2007) from
the evolutionary tracks of Schaller et al. (1992). The model input
parameters are summarized in Table 1.

The value of $\eta$ depends on the efficiency of mixing of
 ejecta material into cold compressed gas that eventually
 becomes molecular material.
The interface of the ejecta and the shocked ambient medium
 is expected to be turbulent due to various instabilities.
Especially, when the elapsed time of supernova expansion becomes
comparable
 to the cooling timescale of post-shock gas, thermal instability makes
 the interface highly turbulent.
This process is clearly shown in hydrodynamical simulations of the
propagation of a shock wave or of a shocked layer  into warm neutral
medium, which results in the creation of cold turbulent clumps
embedded in warm neutral medium via thermal instability (Koyama \&
Inutsuka 2002; Audit \& Hennebelle 2005). The spatial scale of the
smallest turbulent eddy is probably comparable to
  the characteristic size ($\lambda_{\rm C}$) of the smallest cold
  clump that is on the order of the critical length scale ($<0.01$ pc)
  of thermal instability.
  The characteristic mass ($M_{\rm C}$) of the smallest cold clumps
  is much smaller than the solar mass and can be given
  by the following equation:
%The spatial scale of the turbulent eddy is comparable to
% the characteristic size ($\lambda_{\rm C}$) of the cold
% clump which is very small and essentially on the order of
 %the critical length scale ($<0.01$ pc) of thermal instability.
 %The characteristic mass ($M_{\rm C}$) of the cold clumps is
 %much smaller than the solar mass and can be given
 %by the following equation:
 \begin{equation}
      M_{\rm C} \equiv \rho _{\rm C} \lambda_{\rm C}^3
                 \sim 10^{-5} M_{\sun}
                 \left( \frac{\rho_{\rm C}}{10^{-21}~\rm{g~cm^{-3}}} \right)
                 \left( \frac{\lambda_{\rm C}}{0.01~{\rm pc}} \right)^3,
 \end{equation}
 where $\rho_{\rm C}$ is the gas density of cold clumps.
Thus, the mixing of the metal-rich ejecta and the ambient medium
 should be very efficient on this small mass scale.
Therefore, we expect the efficiency of mixing to be very high, and
 use $\eta=1$ hereafter, in line with the value of $\eta$
 adopted by Looney et al. (2006) in the case of a disk and a starless core.
 Given that in our model, the supernovae from OB1 by definition face MC2 (see Fig. 1), we expect $f$ to be close to 0.5. We
conservatively assume that only one tenth of the supernova ejecta
contributes to the sweeping-up of atomic gas and the adduction of
$^{60}$Fe into the new molecular cloud, and therefore adopt $f=0.1$.
%This low
%value of $f$ takes into account the possibility for $^{60}$Fe atoms
%to escape the Galactic disk within Galactic fountains (e.g. Bally et
%al. 2005).

The evolution of $^{60}$Fe in MC2 is calculated for different sizes
of MC1, i.e. for different values of its number of stars, N$_1$. For
each N$_1$, the calculation is realized about 100 times to account
for the stochastic nature of star formation. A typical example, with
N$_1$ = 5000 stars, is given in Fig. \ref{fig-evolution} where each
thin line represents one realization of the simulation, while the
thick red line is the average of 102 realizations. The average
number of supernovae is 11.8, varying from from 4 to 22. Increasing
the number of realizations does not change the shape of the red
average curve depicted in Fig. \ref{fig-evolution}, nor any of the
calculated properties, indicating that 100 realizations of the IMF
suffice to give a fair account of the $^{60}$Fe abundance expected
in MC2. The positive slope at small times (or large masses) is due
to the rarity of occurrence of very massive stars. The positive
slope at t = 16 Myr is due to the high $^{60}$Fe yield of the 13
M$_{\sun}$ star compared to stars with neighboring masses (see Table
1). Note that both Woosley \& Weaver (1995) and Limongi \& Chiffi
(1996) find similar high $^{60}$Fe yields for the 13 M$_{\sun}$
model.
% (10.5 $\times$ 10$^{-5}$ and 7.56
%$\times$ 10$^{-5}$ M$_{\sun}$) respectively.

Because MCs need between 10 and 20 Myr to be built by turbulent
convergent flows (e.g. V\'{a}zquez-Semadeni et al. 2007; Hennebelle
et al. 2008), we calculate the average of the $^{60}$Fe abundance
between 10 and 20 Myr, $\hat{ M}_{\rm MC2}$ ($^{60}\rm Fe$). In our
typical case (N$_1$ = 5000 stars), $\hat{ M}_{\rm MC2}$ ($^{60}\rm
Fe$) = (3.2 $\pm$ 2.0) $\times$ 10$^{-6}$ M$_\sun$, where the
uncertainty corresponds to one standard deviation (Fig.
\ref{fig-var-w}). We find that the value of $\hat{ M}_{\rm MC2}$
($^{60}\rm Fe$) scales linearly with N$_1$, the number of stars in
MC1.

From Fig. \ref{fig-evolution} and \ref{fig-var-w}, it is clear that
second generation MCs are expected to contain a significant amount
of $^{60}$Fe due to contamination by supernovae of a first
generation of stars. It remains to compare that amount of $^{60}$Fe
contained in MC2 to its abundance in the ESS. Note that because the
observed collapse timescales of cores (a few 10$^5$ yr, Onishi et
al. 2002) are far shorter than the $^{60}$Fe half-life, there is no
need for an extra decay term between the molecular cloud stage
(assumed to start 10 to 20 Myr after the onset of star formation in
MC1) and the disk stage, implying that the abundance in
protoplanetary disks is identical to that of MC2 from which they
form.

\subsection{Comparison with the solar system} \label{sec-fe}

The initial abundance of $^{60}$Fe in the solar system is not
precisely known (Mostefaoui et al. 2005; Gounelle \& Meibom 2008).
The most recent and precise studies failed to detect an isochron and
placed upper limits of 6 $\times$ 10$^{-7}$ and 1 $\times$ 10$^{-7}$
respectively for the initial $^{60}$Fe/$^{56}$Fe ratio (Dauphas et
al. 2008; Regelous et al. 2008). Adopting a conservative initial
ratio $^{60}$Fe/$^{56}$Fe = 3 $\times$ 10$^{-7}$, it is estimated
that the molecular cloud progenitor of our solar system had an
$^{60}$Fe concentration of [$^{60}$Fe]$_{\rm SS}$ = 4 $\times$
10$^{-10}$ M$_{\sun}$ per unit of solar mass, assuming a
$^{56}$Fe/$^1$H ratio of 3.2 $\times$ 10$^{-5}$, and a metallicity
of 0.7 (Lodders 2003).

%Different estimates determined by Multi Collector-Inductively
%Coupled Mass Spectrometry (MC-ICPMS) or Secondary Ionization Mass
%Spectrometry (SIMS) vary between 5 $\times$ 10$^{-8}$ and 1 $\times$
%10$^{-6}$ for the initial $^{60}$Fe/$^{56}$Fe ratio (see the
%discussion in Gounelle \& Meibom (2008)). Measurements performed
%with MC-ICPMS failed so far to detect an isochron indicative of the
%past presence of $^{60}$Fe (e.g. Dauphas et al. 2008). The most
%precise work performed to date on primitive carbonaceous chondrites
%by Regelous et al. (2008) constrained the initial
%^{60}$Fe/$^{56}$Fe ratio to be lower than 1 $\times$ 10$^{-7}$,
%suggesting unresolved analytical difficulties.

Our model can account for the $^{60}$Fe abundance in the ESS
provided it formed in a molecular cloud with a mass $M_{\rm
MC2}=\hat{ M}_{\rm MC2}$($^{60}\rm Fe$)/[$^{60}$Fe]$_{\rm SS}$  =
(0.80 $\pm$ 0.50) $\times$ 10$^4$ M$_{\sun}$. Our typical case,
N$_1$ = 5000 stars, corresponds to the estimated number of stars in
the UCL-LCC association which formed $\sim$ 12 Myr before the Upper
Sco association (de Geus 1992), within the 10-20 Myr interval
defined above. If we take 2350 M$_{\sun}$ as the stellar content of
Upper Sco (de Geus 1992) and a molecular gas mass of (0.8 $\pm$ 0.5)
$\times$ 10$^4$ M$_{\sun}$, we obtain a star formation efficiency of
29$^{+50}_{-11}$ \%. This star formation efficiency is in line with
the observed star formation efficiencies (5-30 \%) of nearby
star-forming regions (Lada \& Lada 2003), implying that if Upper Sco
molecular gas was swept-up by the explosions of supernovae from the
UCL-LCC association (Preibisch \& Zinnecker 2007), it is expected to
contain $^{60}$Fe at a concentration similar to that of the ESS.
This indicates that our model is self-consistent and offers a
plausible astrophysical setting for the presence of $^{60}$Fe within
the nascent solar system.

\section{Discussion}

\label{sec-discussion}

\begin{figure}
%\epsscale{1.8}
\plotone{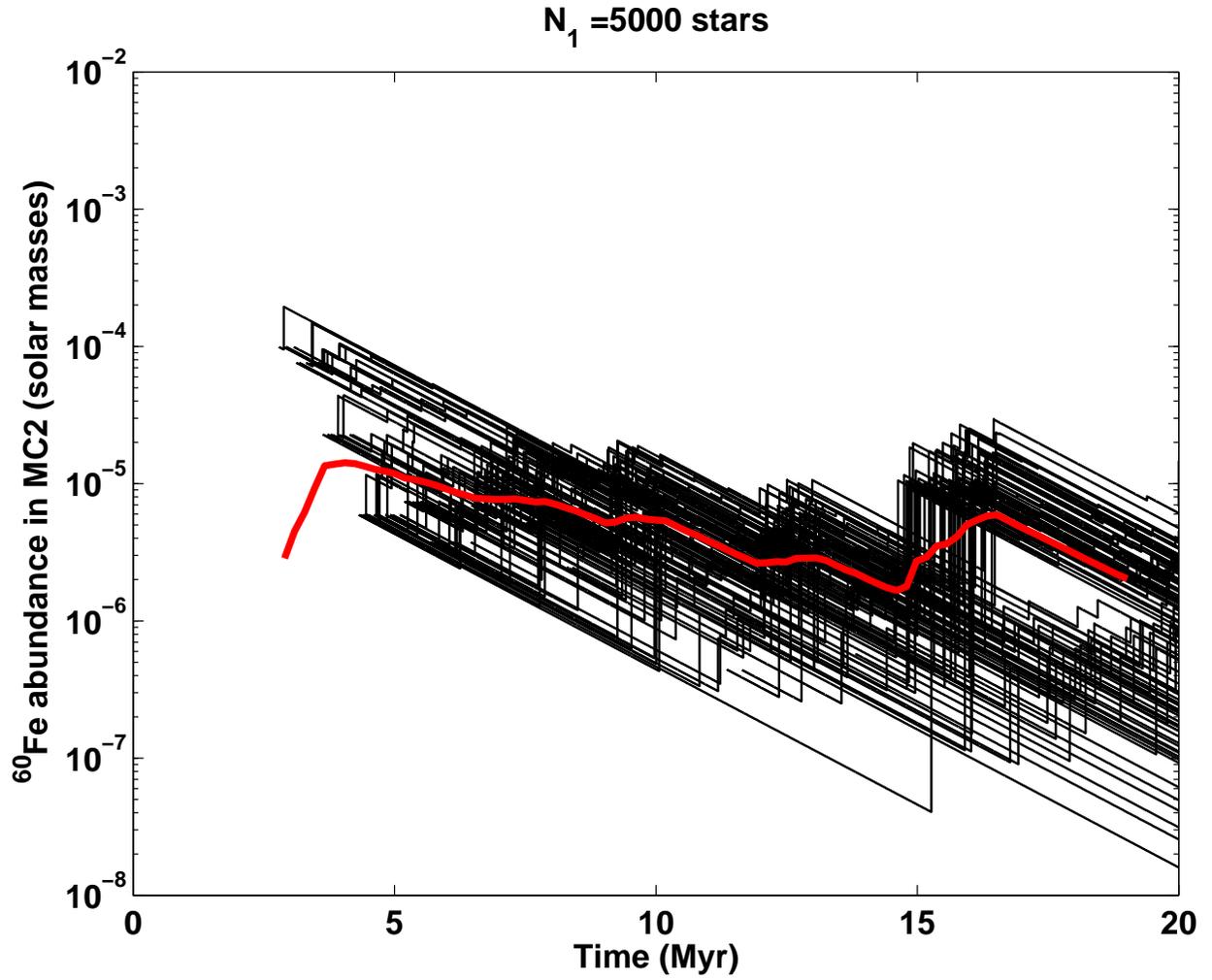} \caption{Time evolution of the $^{60}$Fe abundance
in the MC2 molecular cloud. Each thin line represents one
realization of the simulation, while the thick red line is the
average of 102 realizations.} \label{fig-evolution}
\end{figure}

\begin{figure}
%\epsscale{1.8}
\plotone{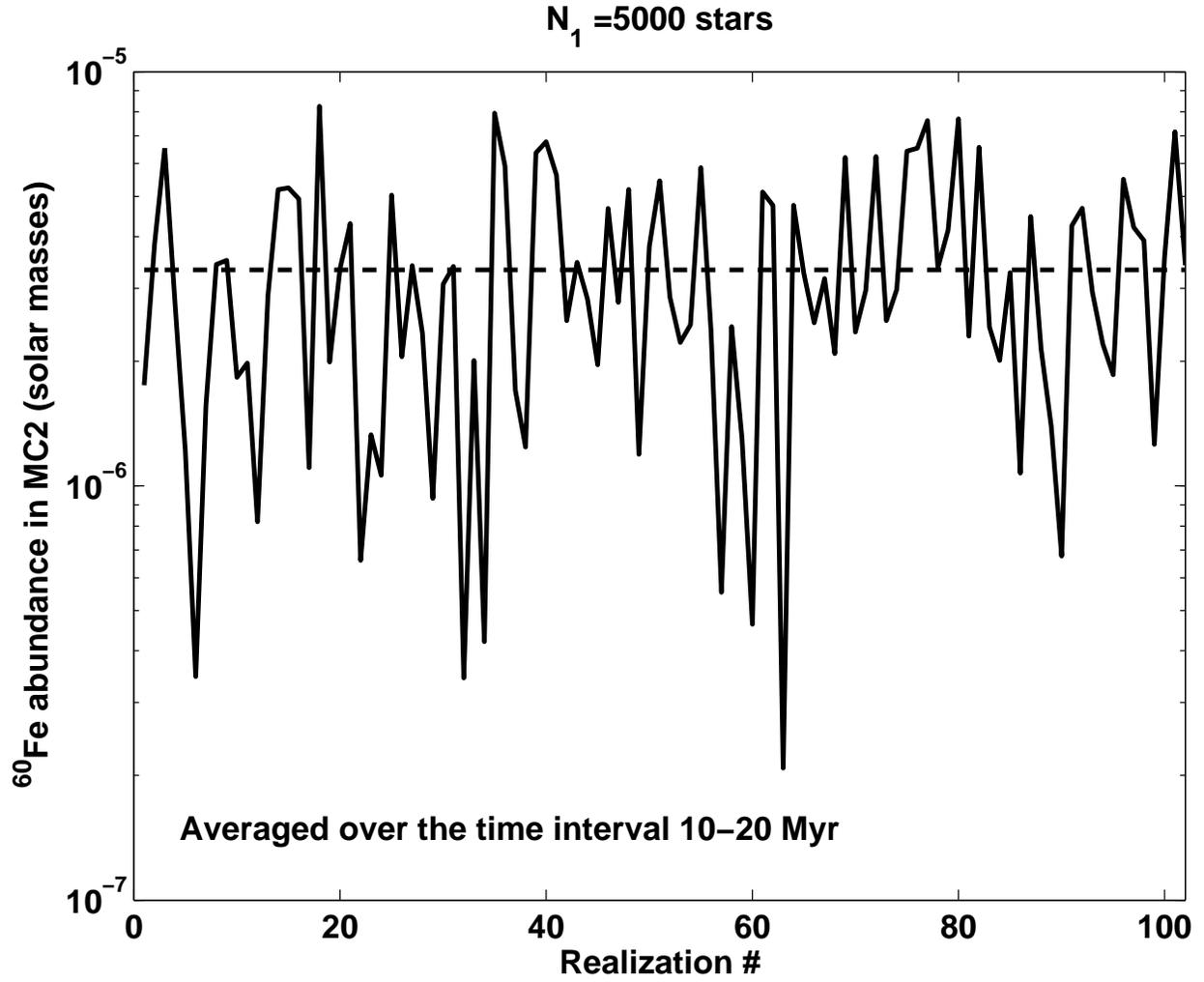} \caption{Average of the $^{60}$Fe abundance in the
MC2 molecular cloud over the time interval 10-20 Myr for different
realizations of the model (100).} \label{fig-var-w}
\end{figure}

It is obvious that given the stochastic nature of star formation and
the variable formation timescales of molecular clouds, a range of
$^{60}$Fe abundance is expected in molecular clouds (Fig.
\ref{fig-evolution} and \ref{fig-var-w}), and therefore in
protoplanetary disks. Some of the input parameters such as $f$ or
$\eta$ could be a factor of a few smaller than the adopted values,
lowering accordingly $\hat{ M}_{\rm MC2}$ ($^{60}\rm Fe$). The
$^{60}$Fe content of the ESS might however be a factor of 3 smaller
than the one we adopted (Regelous et al. 2008). In addition, $f_{\rm
SN}$ could be significantly higher ($f_{\rm SN} = 3 \times 10^{-3}$,
Adams \& Laughlin 2001) than the value we adopted ($f_{\rm SN} = 2.1
\times 10^{-3}$), resulting in a higher value of $^{60}$Fe in MC2
due to a larger number of SNe in MC1. Finally, any increase of the
half-life of $^{60}$ Fe (G\"{u}nter Korschinek, personal
communication) would increase (exponentially) $\hat{ M}_{\rm MC2}$
($^{60}\rm Fe$). The point of the calculations above is to show that
for typical numbers ($^{60}$Fe yields, molecular clouds masses and
formation timescales, star formation efficiency...), the estimated
ESS abundance of $^{60}$Fe can be reproduced in the context of a
reasonable astrophysical model.

%The $^{60}$Fe yields of core-collapse supernovae themselves vary greatly
%depending on the stellar mass but also on the stellar model and
%nuclear physics used (Woosley \& Weaver 1995; Rauscher et al. 2002;
%Limongi \& Chiffi 2006). Despite this variation, it is remarkable
%that for an average $^{60}$Fe yield, molecular clouds/dense shells
%with realistic masses and formed by the mechanisms described above
%(\S \ref{sec-quant}) contain $^{60}$Fe at the solar system
%abundance.

%Though a modern version of the pioneering work of Cameron \& Truran
%(1977),
Our model differs from previous models on several important points.
First, $^{60}$Fe is delivered to a molecular cloud by a diversity of
supernov{\it ae} rather than by a single supernova. Second, the mass
of the receiving phase ($\sim$ 10$^4$ M$_{\sun}$) is orders of
magnitude larger than for the disk and the core model (0.013 and 1
M$_{\sun}$ respectively, e.g. Looney et al. 2006).
%The large scale
%injection of $^{60}$Fe puts our model together with models exploring
%galactic background origin of SLRs (e.g. Meyer \& Clayton 2000).
%The larger size of the receiving region implies that the
%dilution factor $f$ is many orders of magnitude larger in our model
%compared to other models, increasing {\it in fine} the concentration
%in the receiving region.
Third, $^{60}$Fe is not injected into a dense phase ($n_{\rm H_2}
\sim$ 10$^5$ cm$^{-3}$ and $\sim$ 10$^{14}$ cm$^{-3}$ for the core
and disk respectively) isolated from the rest of the ISM, but is
delivered into a relatively diffuse ISM phase interacting with other
ISM components leading to high mixing efficiency (see \S
\ref{sec-quant}). Fourth, the $^{60}$Fe-producing supernovae belong
to previous generations of massive stars rather than to the same
generation of stars. Fifth, the supernovae shock waves do not
trigger the collapse of a pre-existing molecular cloud core, but
rather contribute to build on a timescale of 10-20 Myr a new, second
generation MC.
%Sixth, while all other models look at the pollution
%of {\it one} disk/core by a supernova ejecta, we propose the
%pollution took place at the molecular cloud stage, meaning that {\it
%all} protoplanetary disks formed in that cloud will be enriched in
%$^{60}$Fe.
Finally, this new model takes quantitatively into account
the stochasticity of the star-forming process (though this approach
has been used in a different context, Cervi{\~n}o et al. 2000).

% The Cameron \& Truran (1977) model does
%not investigate the origin of $^{60}$Fe. It considers a small
%interstellar cloud (10$^3$ M$_{\sun}$) in which no star massive
%enough to go supernova is expected (Williams \& Gaidos 2007).

%If we take at t = 10

The paradigm for MC formation described in \S \ref{sec-sketch} is
however not universally accepted. An alternative or complementary
view is that gravitational instabilities represent the main driver
for MC formation rather than large scale  convergent flows (e.g.
Hennebelle et al. 2008). However, independently of the main driver
for MC formation, it remains observationally true that most stars
form in Giant Molecular Clouds (GMCs). In such a context, MC1 and
MC2 would represent two different regions of the same GMC which have
evolved at different paces (e.g. Fellhauer \& Kroupa 2005). The
aforementioned observation that OB associations are made of
subgroups of different ages can indeed be interpreted as an evidence
for sequential star formation in GMCs (Elmegreen \& Lada 1977),
preserving the essence of our model.

The strength of our model is that the $^{60}$Fe content in the ESS
is easily reproduced within a plausible, if not common,
astrophysical setting, unlike the disk or core models (\S 1), and
that all solar systems (within MC2 or the younger region of a GMC)
will receive $^{60}$Fe instead of a few disks or cores. Therefore,
even if a significant fraction of stars (50 \%) form in MC1, the
overall probability for $^{60}$Fe MC inheritance is far higher than
that of injection into a disk or a core. In that respect, though
there is not {\it one} typical solar system, there is no need to
call for an unlikely astrophysical setting for our solar system
formation.

An important and so far unresolved problem associated with all
models based on supernova ejecta, is the over production of
$^{53}$Mn relatively to $^{60}$Fe and their inferred ratio in the
early solar system (Wasserburg et al. 2006). In the current
supernova 1D models, $^{53}$Mn and $^{60}$Fe are produced deep in
the supernova interior, together with $^{56}$Ni ($T_{1/2} = 6 $
days). Nickel-56 is known to make it to the surface (Arnett et al.
1989), ruling out fallback as a solution to the $^{53}$Mn
overproduction problem (e.g. Meyer \& Clayton 2000). Our model does
not offer a solution to this general problem, but it is not
inconceivable that a solution might come from new developments in
supernovae nucleosynthetic models, many aspects of which are not
fully understood (Woosley \& Heger 2007; Magkotsios et al.2008).

% and $^{56}$Ni ($T_{1/2} = 6 $ days)

Though $^{26}$Al can be produced by energetic particles irradiation
(Lee et al. 1998), it might be difficult for irradiation to account
for the entire ESS inventory (Duprat \& Tatischeff 2007; Fitoussi et
al. 2008). The ESS $\rm ^{26}Al / ^{60}{\rm Fe}$ mass ratio was
$\sim$ 8, adopting $^{60}$Fe/$^{56}$Fe = 3 $\times$ 10$^{-7}$,
$^{26}$Al/$^{27}$Al = 5 $\times$ 10$^{-5}$ and an
$^{27}$Al/$^{56}$Fe ratio of 0.11 (Lodders 2003). Given the strong
heterogeneity in the $^{26}$Al distribution (Diehl et al. 2006),
this compares relatively well with the observed ISM $\rm ^{26}Al /
^{60}{\rm Fe}$ mass ratio of $\sim$ 3 (Wang et al. 2007). This
suggests that together with $^{60}$Fe, a substantial amount of
$^{26}$Al could also have been inherited from the progenitor
molecular cloud.

\acknowledgments M. Serrano is thanked for producing Fig. 1. The
reviewers significantly helped to improve the paper. This study was
funded by the PNP and the France-{\'E}tats-Unis fund from CNRS, the
ANR and the european grant ORIGINS [MRTN-CT-2006-035519].

\clearpage

\clearpage

\clearpage

\clearpage
\begin{deluxetable}{ccc}

\tablewidth{0pt}

\tablecaption{Model input data}

\tablehead{\colhead{$M$ (M$_{\sun}$)}&\colhead{ $t_{\rm SN}$(Myr)} &
\colhead{ $ Y_{\rm SN}(^{60}{\rm Fe})$ (M$_{\sun}$)}  }

\startdata
11 & 20.8 & 5.25E-6  \\
12 & 17.8 & 3.62E-6 \\
13 & 15.5 & 9.03E-5  \\
14 & 13.8 & 5.72E-6 \\
15 & 12.5 & 3.31E-5  \\
16 & 11.4 & 4.39E-6 \\
17 & 10.6 & 7.96E-6  \\
18 & 9.9 & 2.54E-5 \\
19 & 9.2 & 7.83E-5  \\
20 & 8.8 & 2.09E-5 \\
21 & 8.3 & 2.45E-5  \\
22 & 7.9 & 5.19E-5 \\
25 & 7.0 & 6.96E-5  \\
30 & 6.0 & 3.75E-5 \\
35 & 5.3 & 7.37E-5  \\
40 & 4.9 & 5.93E-5 \\
60 & 3.9 & 2.27E-4  \\
80 & 3.4 & 7.55E-4 \\
120 & 3.9 & 9.93E-4  \\
\enddata
\tablecomments{$M$ is the stellar mass, $t_{\rm SN}$ is the stellar
lifetime and $ Y_{\rm SN}(^{60}{\rm Fe})$ the $^{60}$Fe yield. The
stellar lifetime is calculated using the formula ${\rm log}({
t}_{\rm SN})={1.4}/{({\rm log } M_{\rm })^{1.5}}$ (Schaller et al.
1992; Williams \& Gaidos 2007). Iron-60 yields are the average of
the yields modelized by Woosley \& Weaver (1995), Rauscher et al.
(2002) and Limongi \& Chiffi (2006).}
\end{deluxetable}

\end{document}